\documentclass{article}

\begin{document}
\def\kpipi{$D^+\rightarrow K^-\pi^+\pi^+$\ }
\def\kpp{$K\pi\pi$\ }
\def\dskkpi{$D^+_s\rightarrow K^-K^+\pi^+$\ }
\def\dskstk{$D^+_s\rightarrow \bar K^*K^+$\ }
\def\dsphipi{$D^+_s\rightarrow \phi\pi^+$\ }
\def\ds3pi{$D_s^+\rightarrow \pi^-\pi^+\pi^+$\ }
\def\d3pi{$D^+\rightarrow \pi^-\pi^+\pi^+$\ }
\title{Light Mesons and Charm Decays: New Results from E791}
\author{Carla G\"obel $^\dag$       \\
{\em Instituto de F\'\i sica, F. Ing., Univ. Rep\'ublica, Montevideo, Uruguay} \\
      \\
{$^\dag$ {\small on behalf of the E791 Collaboration}}
}
\date{}
\maketitle
\baselineskip=11.6pt
\begin{abstract}
We will discuss how the decays of charm mesons can be used to study light 
mesons spectroscopy, by presenting recent results of Dalitz plot 
analyses using data from Fermilab experiment E791. Emphasis will be on scalar 
mesons, which are found to have large contribution to the $D$ decays studied. 
In addition to the usual extraction of decay fractions and relative phases 
of the intermediate amplitudes, the Dalitz plot technique is used to measure masses
and widths of scalar resonances.
From the $D_s$ decay, we obtain masses and widths of $f_0(980)$ and $f_0(1370)$.
We find evidence for a light and broad scalar resonance, the $\sigma$ meson, 
in $D^+\to\pi^-\pi^+\pi^+$ decay. Preliminary studies also show evidence for 
a light and broad resonance, the $\kappa$ meson, in $D^+\to K^-\pi^+\pi^+$ 
decay. These results illustrate the potential of charm decays as a
laboratory for the study of light mesons.
\end{abstract}
\baselineskip=14pt
%

\section{Introduction}
%
%

The decays of $D$ mesons can be seen as a new environment for the study of light 
meson physics, due to the high quality of the current samples, the fact that the initial
state is always well defined (defined mass, $0^-$ state), and usually with a small 
non-resonant contribution. Particularly, these decays can shed light on the scalar 
mesons, some of them long-standing sources of uncertainty.

Here we present results for the Dalitz plot analyses of the decays \ds3pi\cite{ds3pi} and
\d3pi\cite{d3pi}, and also preliminary results for the decay $D^+\to K^-\pi^+\pi^+$, 
using data from the 
Fermilab E791 experiment\footnote{E791 ran in 91-92 with a 500 GeV/c 
$\pi^-$ beam. See \cite{e791refs}.}. 

In the \ds3pi decay, previous measurements showed the dominance of scalar resonant 
states\cite{e691-3pi,e687-3pi}. For instance, the decay $f_0(980)\pi^+$ seems to 
contribute most. Nevertheless, the nature of the $f_0(980)$ is still a puzzle, and 
in particular its width is poorly measured\cite{pdg}. From
the E791 Dalitz plot analysis of the \ds3pi decays, we are able to obtain 
measurements for the mass and width of this state, as well as for the $f_0(1370)$.
For the \d3pi decay, we are not able to get a good description of the data using
the same intermediate states as for \ds3pi. We find strong evidence for a new light and
broad scalar resonance, the $\sigma(500)$, and measure its mass and width. 
The $D^+\to\sigma\pi^+$ channel accounts for about half of the 
total \d3pi decay rate. Other experiments have presented
inconsistent evidence for low-mass $\pi\pi$ resonances in partial wave analyses 
\cite{sigma-exp}, resulting in ambiguous results for the characteristics of such particles
\cite{pdg,torn}.

For the \kpipi decay, we present preliminary results from a high statistics sample
(almost 23,000 decays). Like previous results for this channel\cite{e691-kpipi,e687-kpipi}
we find that the non-resonant decay is dominant, which is unusual in $D$ decays, 
but also find important discrepancies between fit model and data. By including
a new scalar state, with unconstrained mass and width, we get a much better description
of the data. This resonance appears as light and broad according to the fit. 
Present discussions about the existence of such a state, referred to as the $\kappa$ in the 
literature, are very controversial\cite{refkappa}.

\section{$D^+, D_s^+\to \pi^-\pi^+\pi^+$ Dalitz plot analyses}

\begin{figure}[t]
\vspace{4.7cm}
\includegraphics{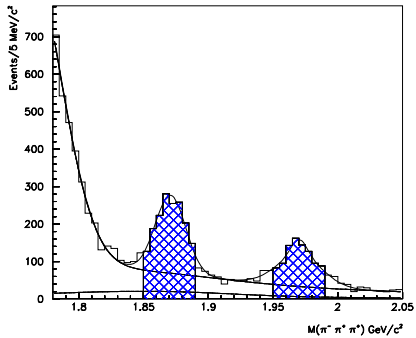}
\caption{\it The $\pi^-\pi^+\pi^+$ invariant mass spectrum. The dotted line
represents the $D^0\to K^-\pi^+$ plus $D_s^+\to \eta'\pi^+$ reflections and 
the dashed line is the total background. Events used for the Dalitz analyses
are in the hatched areas.
\label{m3pi} }
\end{figure}

In Fig.~\ref{m3pi} we show the $\pi^-\pi^+\pi^+$ invariant mass distribution for
the sample collected by E791 after reconstruction and selection criteria \cite{ds3pi,d3pi}. 
Besides combinatorial background, reflections from the decays \kpipi, $D^+\to K^-\pi^+$ 
(plus one extra track) and $D_s^+\to \eta'\pi^+,
~\eta'\to\rho^0(770)\gamma$ are all taken into account.
The hatched regions in Fig.~\ref{m3pi} show the samples used for the
Dalitz analyses. There are 937 and 1686 candidate events for $D_s^+$ and $D^+$ respectively,
with a signal to background ratio of about 2:1. 

\begin{figure}[t]
\vspace{5.0cm}
\includegraphics{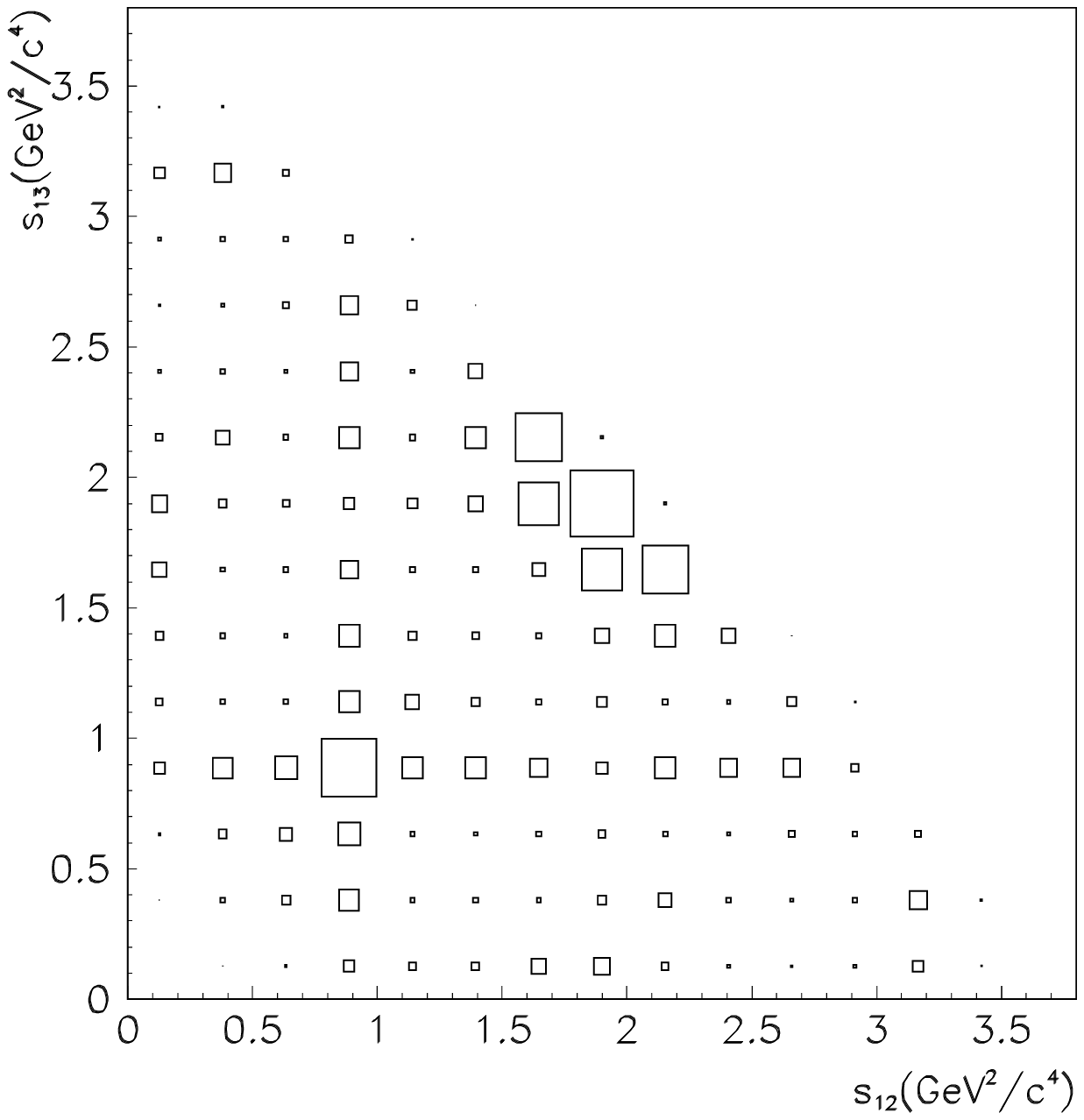} 
\includegraphics{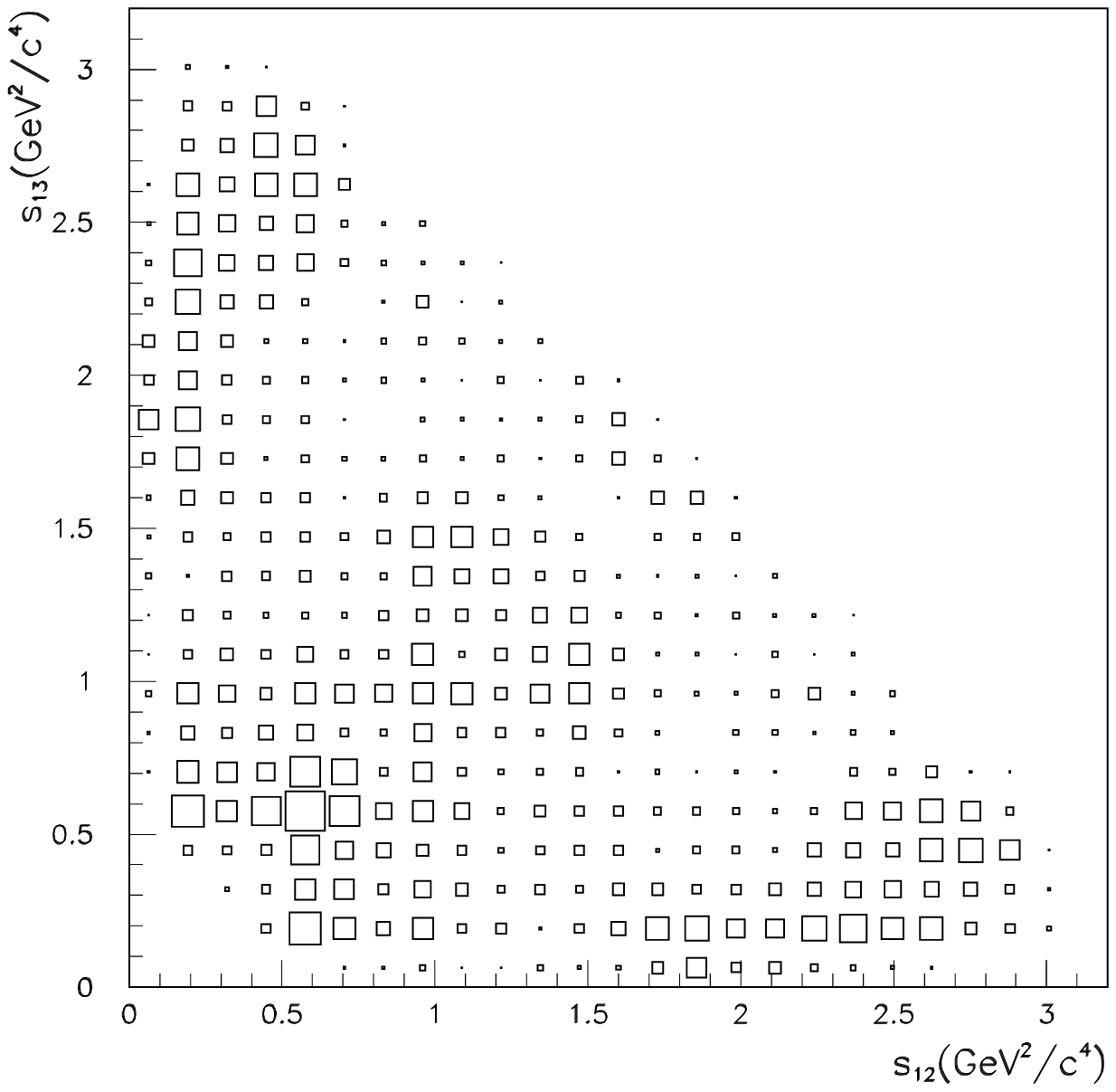} 
\hspace*{3cm} (a) \hspace*{4cm} (b)
\caption{\it {\rm (a)} The $D_s^+ \to \pi^- \pi^+ \pi^+$ Dalitz plot and
{\rm (b)} the \d3pi Dalitz plot. Since there are two identical particles, 
the plots are symmetrized.
\label{dalitz3pi} }
\end{figure}

The Dalitz plot corresponding to \ds3pi events is shown in Fig.~\ref{dalitz3pi}(a).
The narrow horizontal and vertical bands at $s_{12}\equiv m^2(\pi^-_1 \pi^+_2)$ 
and $s_{13}\equiv m^2(\pi^-_1 \pi^+_3)$ just below 1 GeV$^2/c^4$
correspond to the $f_0(980) \pi^+$ state.
At the upper edge of the diagonal, there is another concentration of events
centered at $s_{12} \simeq  s_{13} \simeq $1.8 GeV$^2/c^4$, corresponding to the
$f_2(1270) \pi^+$, $f_0(1370) \pi^+$, and $\rho^0(1450) \pi^+$ contributions.
From the \d3pi Dalitz plot (Fig.~\ref{dalitz3pi}(b)) we see clearly
the bands corresponding to the $\rho^0(770)\pi^+$ and $f_0(980)\pi^+$ channels,
and an excess of events at low $\pi^- \pi^+$ invariant-mass squared.

To study the resonant structure of these decays, unbinned maximum likelihood fits are
applied to the samples. The Dalitz plot distributions are fitted to a sum of
signal and background probability distribution functions (PDFs). For each candidate event,
the signal PDF is written as the square of the total physical amplitude, which is
constructed from a coherent sum of Lorentz invariant amplitudes corresponding to  
the various resonant channels plus the non-resonant decay. The signal PDF is weighted
by the acceptance across the Dalitz plot (obtained by Monte Carlo (MC)) and by 
the level of signal to background for each event, as given by the line shape of 
Fig.~\ref{m3pi}. 

The resonant channels we include in the fits are $\rho^0(770) \pi^+$, $f_0(980) \pi^+$, 
$f_2(1270) \pi^+$, $f_0(1370) \pi^+$, and $\rho^0(1450) \pi^+$.
We assume the non-resonant amplitude to be uniform across the Dalitz plot.
Each resonant amplitude, except that  for the $f_0(980)$, is
parameterized as a product of form factors, a relativistic Breit-Wigner (B-W)
function, and an angular momentum amplitude which depends on the spin of the 
resonance, 
\begin{equation}
{\cal A}_n = \frac {F_D\  F_R\ {\cal M}_n^{(J)}} {m_{12}^2 - m_0^2 + im_0\Gamma(m_{12})}
\ ,\,
\Gamma(m_{12}) = \Gamma_0 \frac{m_0}{m_{12}} \left(\frac{p^*}{p^*_0}
\right)^{2J+1} \frac{F_R^2(p^*)}{F_R^2(p^*_0)}\ .
\label{ampl}
\end{equation}

Above, $m_{12}$ is the $\pi^+\pi^-$ invariant mass for the 
candidate resonance. Since there are two like-charge pions, each signal 
amplitude is Bose-symmetrized,  
${\cal A}_n =  {\cal A}_n[({\bf 12}){\bf3}] + {\cal A}_n[({\bf 13}){\bf 2}]$.
The quantities $F_D$ and  $F_R$ are the Blatt-Weisskopf damping factors
\cite{blatt} respectively for the $D$ and the resonance\footnote{For 
three pions analyses, the effective radii for $F_D$ and $F_R$ are set to 3.0GeV$^{-1}$;
for the \kpipi analysis of section \ref{seckpipi}, the radii taken for $F_D$ and $F_R$
are respectively 3.0GeV$^{-1}$ and 1.5GeV$^{-1}$.},
$p^*$ is the pion momentum in the resonance rest frame at mass $m_{12}\
(p^*_0=p^*(m_0))$. ${\cal M}_n^{(J)}$ describes the angular distribution due to the 
spin $J$ of the resonance. 

For the $f_0(980) \pi^+$ we use a coupled-channel B-W function, 
following the parameterization of the WA76 Collaboration\cite{wa76},
\begin{equation}
BW_{f_0(980)} = {1 \over {m_{\pi\pi}^2 - m^2_0 + im_0(\Gamma_{\pi}+\Gamma_K)}}\ ,
\end{equation}
\begin{equation}
\Gamma_{\pi} = g_{\pi}\sqrt{m_{\pi\pi}^2/ 4 - m_{\pi}^2},~
\Gamma_K = {g_K \over 2} \left( \sqrt{m_{\pi\pi}^2/ 4 - m_{K^+}^2}+
\sqrt{m_{\pi\pi}^2/ 4 - m_{K^0}^2}\right) .
\end{equation}

We multiply each amplitude  by a complex coefficient, $c_n=a_ne^{\delta_n}$.
The fit parameters are the magnitudes, $a_n$, and the phases,
$\delta_n$, which accommodate the final state interactions. 

\subsection{Results for \ds3pi}

The \ds3pi Dalitz plot (Fig.~\ref{dalitz3pi}(a)) is fitted to obtain not only the 
relative contributions (and phases) of the possible sub-channels, but also 
the parameters of the $f_0(980)$ state, $g_{\pi}$, $g_K$, and $m_0$, as well 
as the mass and width of the $f_0(1370)$. That is, they are determined directly 
from the data, as free parameters in the fit. The other resonance 
masses and widths are taken from the PDG\cite{pdg}.
The resulting fractions and phases are shown in Table \ref{tabds3pi}. 

The measured $f_0(980)$ parameters are 
 $m_0 =  977 \pm 3 \pm 2$ MeV/c$^2$, $g_{\pi} =$ 0.09  $\pm$  0.01 $\pm$ 0.01 
and  $g_K =$ 0.02  $\pm$  0.04  $\pm$  0.03.
Our value for $g_{\pi}$ is in very good agreement with OPAL and MARKII results \cite{opal}, 
but WA76 \cite{wa76} found a much larger value,
$g_{\pi} =$ 0.28  $\pm$  0.04. Our value of $g_K$ indicates a small coupling of $f_0(980)$ to
$K\bar K$. The values of the $f_0(980)$ mass and of
$g_{\pi}$, as well as  the magnitudes and phases of the resonant amplitudes, 
are relatively insensitive to the value of $g_K$. Both OPAL and MARKII results
are also insensitive to the value of $g_K$. WA76, on the contrary, measured 
$g_K =$ 0.56  $\pm$  0.18. 

We have also fit the Dalitz plot using for the $f_0(980)$ the same
B-W function as for the other resonances. We find 
$m_0 = 975 \pm 3$ MeV/c$^2$ and $\Gamma_0 =  44 \pm 2 \pm 2$ MeV/c$^2$, and the results
for fractions and phases are indistinguishable.

\begin{figure}[t]
\vspace{4.7cm}
\includegraphics{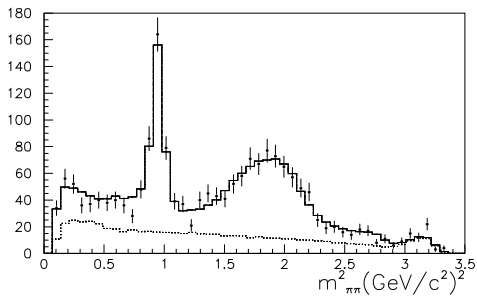}
\caption{\it $s_{12}$ and $s_{13}$ ($m^2_{\pi\pi}$) projections for \ds3pi data (dots) 
and our best fit
(solid). The hashed area corresponds to  background.
\label{proj_ds} }
\end{figure}

A $\chi^2$ distribution is produced from the 
difference in densities for model (from a fast-MC algorithm) and data. 
From the $\chi^2$ and the number of 
degrees of freedom ($\nu$), the confidence level of the model is 35\% \cite{ds3pi}.
A visual comparison between the fit model and the data can be seen in Fig.~\ref{proj_ds}
where we show the sum of the $s_{12}$ and  $s_{13}$ projections for data (points)
and model (solid lines, from fast-MC).

\begin{table}[t]\centering
\caption{\it Dalitz fit results for \ds3pi.}
\vskip 0.1 in
\begin{tabular}{|c|c|c|c|} \hline
Decay Mode         &    Phase($^\circ$) &    Fraction(\%)      \\ \hline 
$f_0(980)\pi^+$    & $0$ (fixed)        & $56.5\pm 4.3\pm 4.7$ \\ 
non-reson.         &~$181\pm 94\pm 51$~ & $ 0.5\pm 1.4\pm 1.7$ \\ 
$\rho^0(770)\pi^+$ & $109\pm 24\pm  5$  & $ 5.8\pm 2.3\pm 3.7$ \\ 
$f_2(1270)\pi^+$   & $133\pm 13\pm 28$  & $19.7\pm 3.3\pm 0.6$ \\ 
$f_0(1370)\pi^+$   & $198\pm 19\pm 27$  & $32.4\pm 7.7\pm 1.9$ \\ 
$\rho^0(1450)\pi^+$& $162\pm 26\pm 17$  & $ 4.4\pm 2.1\pm 0.2$ \\ \hline
\end{tabular}
\label{tabds3pi}
\end{table}

If the $D_s^+ \to \pi^- \pi^+ \pi^+$
decay is dominated by the Cabibbo-favored spectator mechanism, we would
expect final states with a large $s\bar{s}$  content.  
As we can see by the results of Table \ref{tabds3pi}, approximately half of 
the $D_s^+ \to \pi^- \pi^+ \pi^+$ 
rate is produced via  $f_0(980)\pi^+$. If the spectator amplitude is dominant in 
this decay, this would support the interpretation of the $ f_0(980) $
as an $ s \bar s $ state.  On the other hand, the 
large contribution from the  intermediate state  $ f_0 (1370) \pi^+$ 
indicates the  presence of either $ W$-annihilation amplitudes or strong 
rescattering in the final state. In fact, the $ f_0 (1370) \pi^+$ is not observed
in the $D_s^+ \to K^+K^-\pi^+$  final state\cite{e687-kkpi}, pointing to the
$f_0(1370)$ being  a  non-$ s \bar s $  particle, as suggested by the 
naive quark model\cite{pdg}.
There is no evidence in the $D_s^+$ decay for a low-mass
broad scalar particle as seen in the  $D^+$ decay, discussed below.

\subsection{Results for \d3pi}

\begin{figure}[t]
\vspace{4.0cm}
\includegraphics{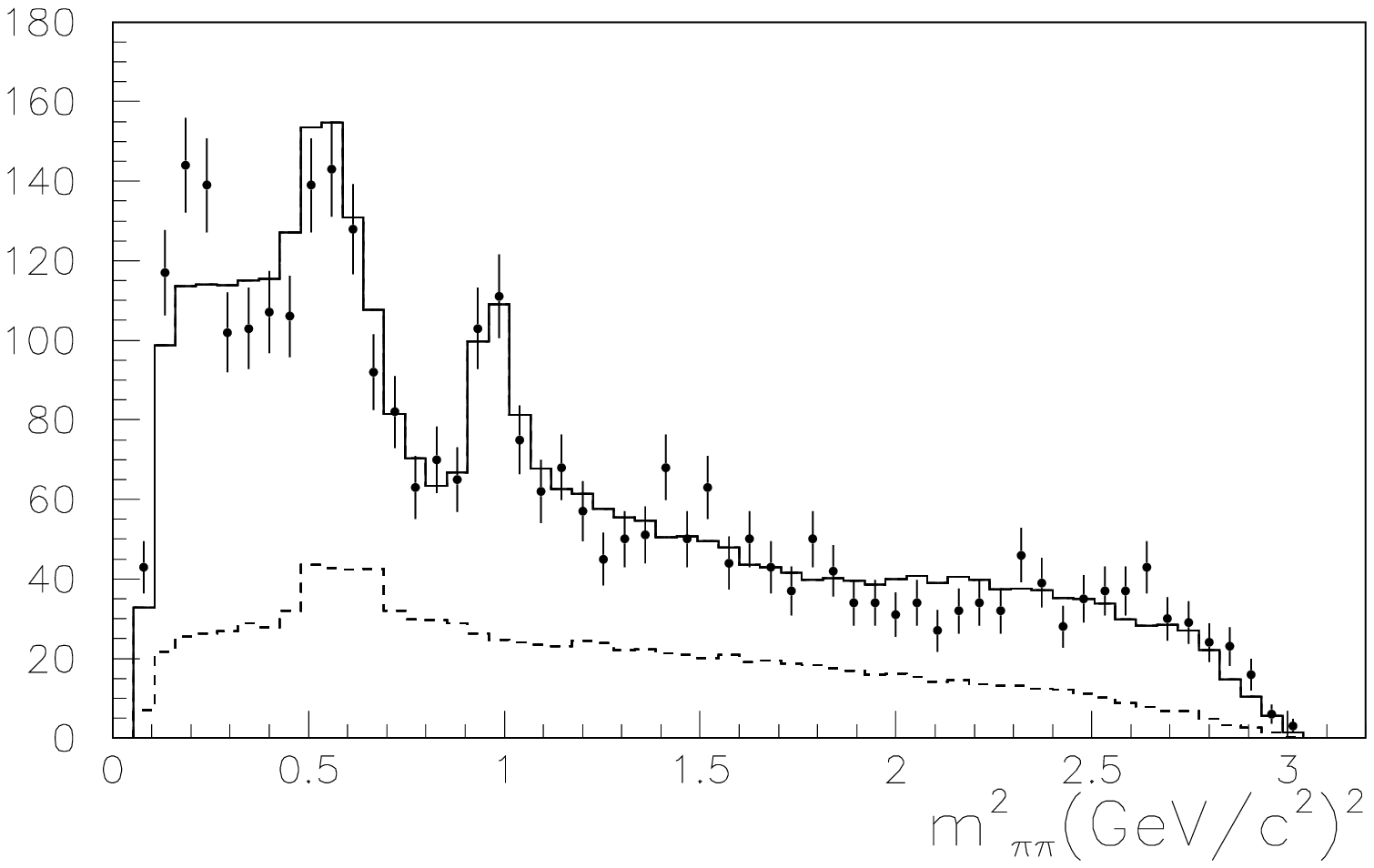}
\includegraphics{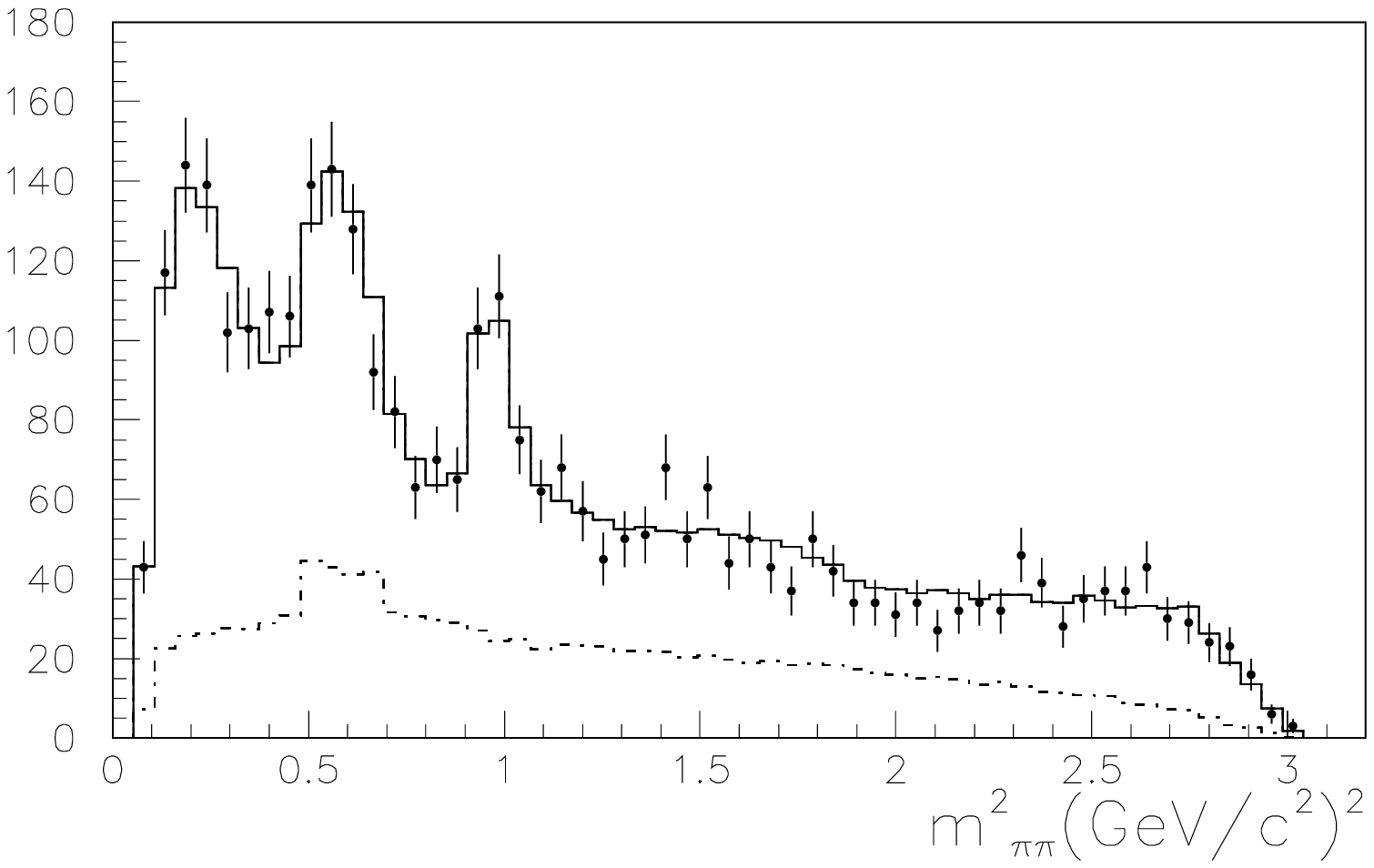}
\hspace*{3cm} (a) \hspace*{5cm} (b)
\caption{\it $s_{12}$ and $s_{13}$ ($m^2_{\pi\pi}$) projections for \d3pi data (dots) 
and our best fit
(solid) for models {\rm (a)} without and {\rm (b)} with $\sigma\pi^+$ amplitude. 
The dashed distribution corresponds to the expected background level.
\label{proj_dp3pi} }
\end{figure}

In a first approach, we try to fit the \d3pi Dalitz plot (Fig.~\ref{dalitz3pi}(b))
using the same amplitudes used for the \ds3pi analysis. We find that, with this model, 
the non-resonant, the $\rho^0(1450)\pi^+$, and the $\rho^0(770)\pi^+$ amplitudes
dominate (results shown on the first column of Table \ref{tabd3pi}), reproducing the
same qualitative features from those reported previously \cite{e691-3pi,e687-3pi}.
However, this model does not describe the data satisfactorily, especially at low 
$\pi^-\pi^+$ mass squared, as can be seen from Fig.~\ref{proj_dp3pi}(a). The $\chi^2/\nu$ 
obtained from the binned Dalitz plot for this model is 1.6, with a confidence level less 
than $10^{-5}$. 
   
To investigate the possibility that another $\pi^-\pi^+$ resonance contributes to the
\d3pi decay, we add an extra scalar resonance amplitude to the signal PDF. 
We allow its mass and width to float as free fit parameters. 

We find that this model improves our fit substantially, converging to values of
mass and width of this scalar resonance, called $\sigma(500)$, of 
$ 478^{+24}_{-23} \pm 17  $ MeV/$c^2$ and 
$ 324^{+42}_{-40}  \pm 21$ MeV/$c^2$, respectively. According to this model, this
amplitude produces the largest decay fraction, as shown in the second column of Table 
\ref{tabd3pi}; the non-resonant amplitude, which is dominant in the model without 
$\sigma\pi^+$, drops substantially. 
This model describes the data much better, as can be seen by the $\pi\pi$ mass squared
projection in Fig.~\ref{proj_dp3pi}(b). The $\chi^2/\nu$ is now 0.9, with a corresponding  
confidence level of 91\%.

To better understand our data, we also fit it with vector, tensor, and toy models
for this extra amplitude, allowing the masses, widths, and relative amplitudes to
float freely. The vector and tensor models test the angular distribution of the
signal. The toy model tests the phase variation expected from a B-W amplitude
by forcing this amplitude to have a constant relative phase but still a B-W shape vs. mass.
All these alternative models fail to describe the data as well as the scalar (regular)
B-W amplitude (see \cite{d3pi}).

\begin{table}[t]\centering
\caption{\it Dalitz fit results for \d3pi. First errors are statistical, second systematics
(only for fit with $\sigma\pi^+$ mode).}
\vskip 0.1 in
\begin{tabular}{|c|c|c|c|c|} \hline  
Decay & \multicolumn{2}{|c|}{Fit without $\sigma\pi^+$} 
      & \multicolumn{2}{|c|}{Fit with $\sigma\pi^+$} \\ \cline{2-5}
Mode               &    Phase($^\circ$)      &    Fraction(\%) 
                   &    Phase($^\circ$)      &    Fraction(\%)  \\ \hline 
$\sigma\pi^+$      &         --               &          -- 
                   & $206\pm 8\pm 5$  & $46.3\pm 9.0\pm 2.1$ \\ 
$\rho^0(770)\pi^+$ & $0$ (fixed)      & $20.8\pm 2.4$ 
                   & $0$ (fixed)      & $33.6\pm 3.2\pm 2.2$ \\ 
non-reson.         & $150\pm 12$      & $38.6\pm 9.7$ 
                   & $57\pm 20\pm 6$  & $ 7.8\pm 6.0\pm 2.7$ \\ 
$f_0(980)\pi^+$    & $152\pm 16$      & $7.4\pm 1.4$  
                   & $165\pm 11\pm 3$ & $ 6.2\pm 1.3\pm 0.4$ \\ 
$f_2(1270)\pi^+$   & $103\pm 16$      & $6.3\pm 1.9$
                   & $57\pm 8\pm 3$   & $19.4\pm 2.5\pm 0.4$ \\ 
$f_0(1370)\pi^+$   & $143\pm 10$      & $10.7\pm 3.1$
                   & $105\pm 18\pm 1$ & $ 2.3\pm 1.5\pm 0.8$ \\ 
$\rho^0(1450)\pi^+$& $ 46\pm 15$      & $22.6\pm 3.7$
                   & $319\pm 39\pm 11$& $ 0.7\pm 0.7\pm 0.3$ \\ \hline
\end{tabular}
\label{tabd3pi}
\end{table}

\section{Preliminary results for \kpipi Dalitz analysis}
\label{seckpipi}

Previous analyses of the Dalitz plot of the decay \kpipi \cite{e691-kpipi,e687-kpipi} 
showed the unusual dominance of the non-resonant decay. The most
recent result comes from E687 \cite{e687-kpipi} with a sample of about 8000 events, where
besides the non-resonant (about 95\% of the decay rate) they find contributions from
the channels $\bar K^*_0(1430)\pi^+$, $\bar K^*(890)\pi^+$, and $\bar K^*(1680)\pi^+$.
However, they don't obtain a good description of the data, arriving to a $\chi^2/\nu$ of
3.

\begin{figure}[t]
\vspace{5.7cm}
\includegraphics{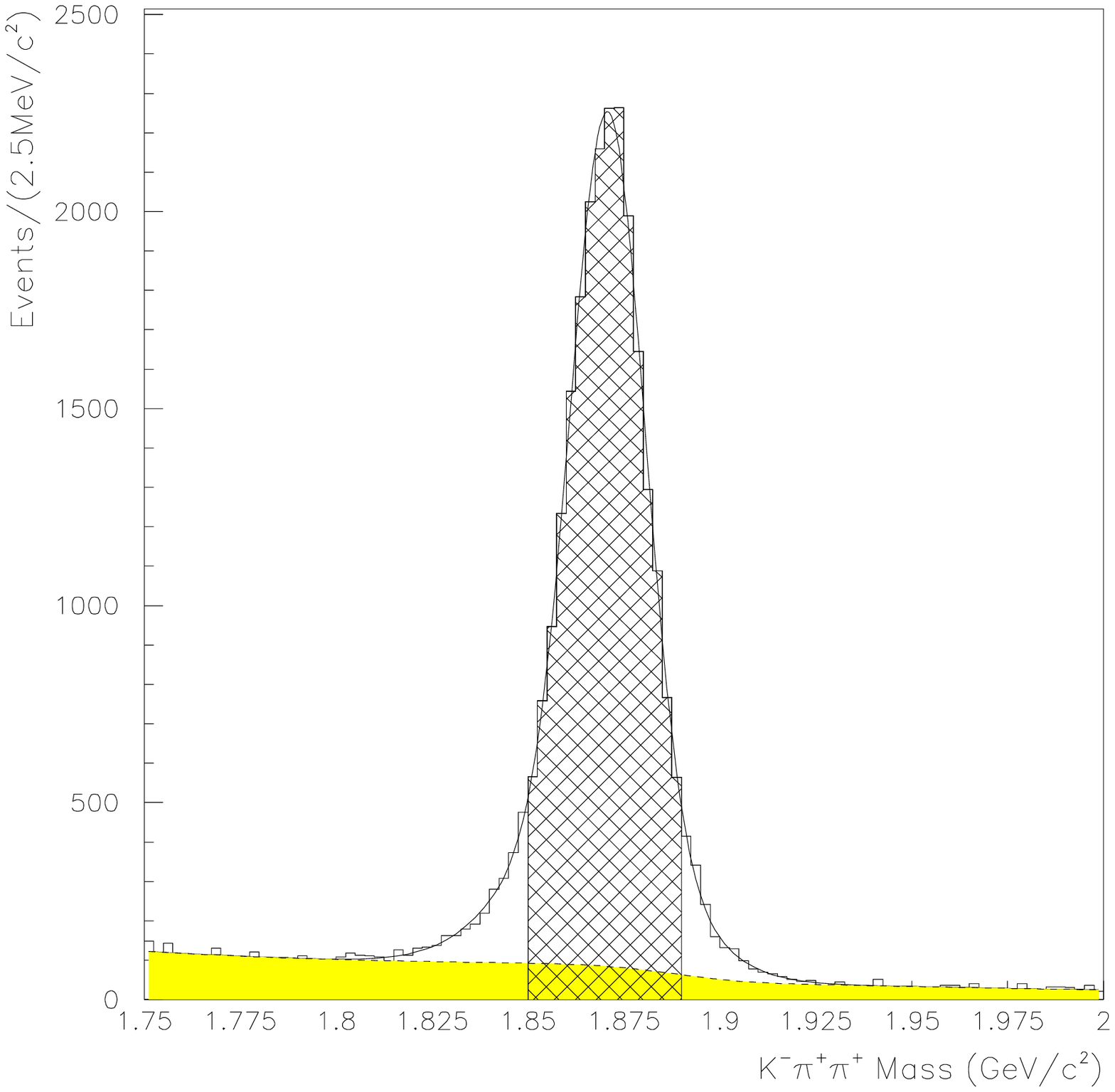}
\includegraphics{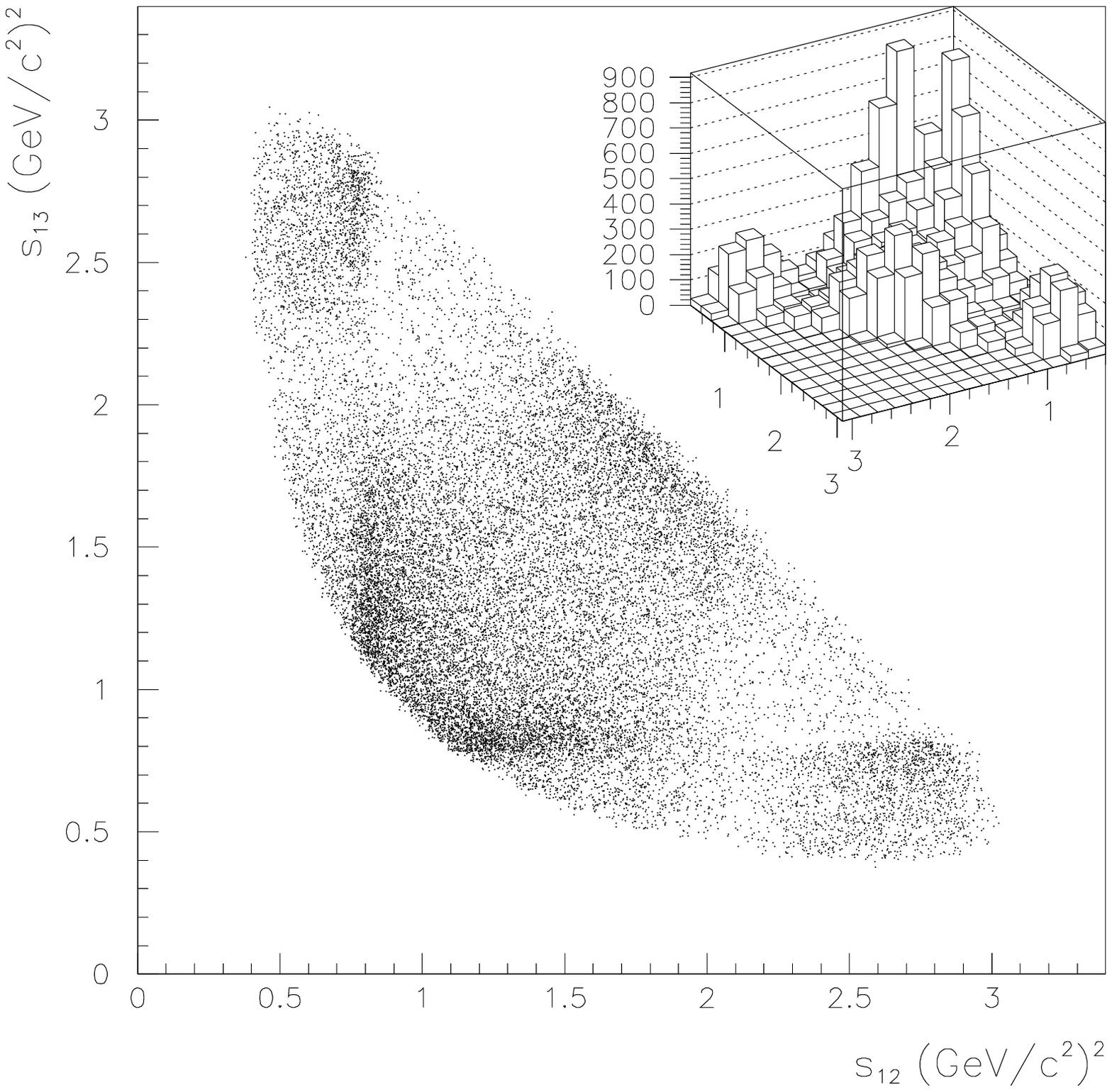}
\hspace*{3cm} (a) \hspace*{5cm} (b)
\caption{\it (a) The \kpp invariant mass spectrum. The shaded area is 
background; (b) Dalitz plot corresponding to the events in the dashed area of (a).
\label{kpipi} }
\end{figure}

In Fig.~\ref{kpipi}(a) we show the $K^-\pi^+\pi^+$ invariant mass distribution for
the sample collected by E791 after reconstruction and selection criteria. Besides
combinatorial background, the other main source of background comes from the reflection 
of the decay $D_s^+\to K^-K^+\pi^+$ (through $\bar K^*K^+$ and $\phi\pi^+$). The total
level of background is shown by the filled area in Fig.~\ref{kpipi}(a). The hashed area 
corresponds to the sample used for the Dalitz analysis. There are 22890 events in this 
sample, where about 6\% correspond to background. 

The Dalitz plot of the signal-region events is shown in Fig.~\ref{kpipi}(b). 
The plot presents a rich structure, where we can observe the clear
bands from $\bar K^*(890)\pi^+$, and an accumulation of events at the upper edge of
the diagonal, due to the heavier resonances.

The formalism used for the \kpipi Dalitz plot fit is essentially the same as for the 
$D^+,D_s^+$ three pions decays\footnote{Besides the differences in the factor 
$F_R$ addressed before, here we define the relativistic Breit-Wigner with a factor of 
(-1) with respect to equation \ref{ampl}, for easier comparison to previous analysis 
\cite{e687-kpipi}.}. The results presented here are all preliminary, in particular no studies
of systematic effects are addressed.

All possible known $K\pi$ resonances \cite{pdg} are included
in the model in a first step. 
We find significant contributions from the same channels 
observed previously\cite{e691-kpipi,e687-kpipi}, but with our higher statistics sample we also 
measure a small but significant contribution from $\bar K^*_2(1430)\pi^+$. The decay
fractions and phases for this model are shown in Table \ref{tabkpipi}, first column.
We confirm the high non-resonant rate (over 100\%); there is an impressive interference
pattern according to this model, reflected by the sum of the fractions being around 150\%.

\begin{figure}[t]
\vspace{7.8cm}
\includegraphics{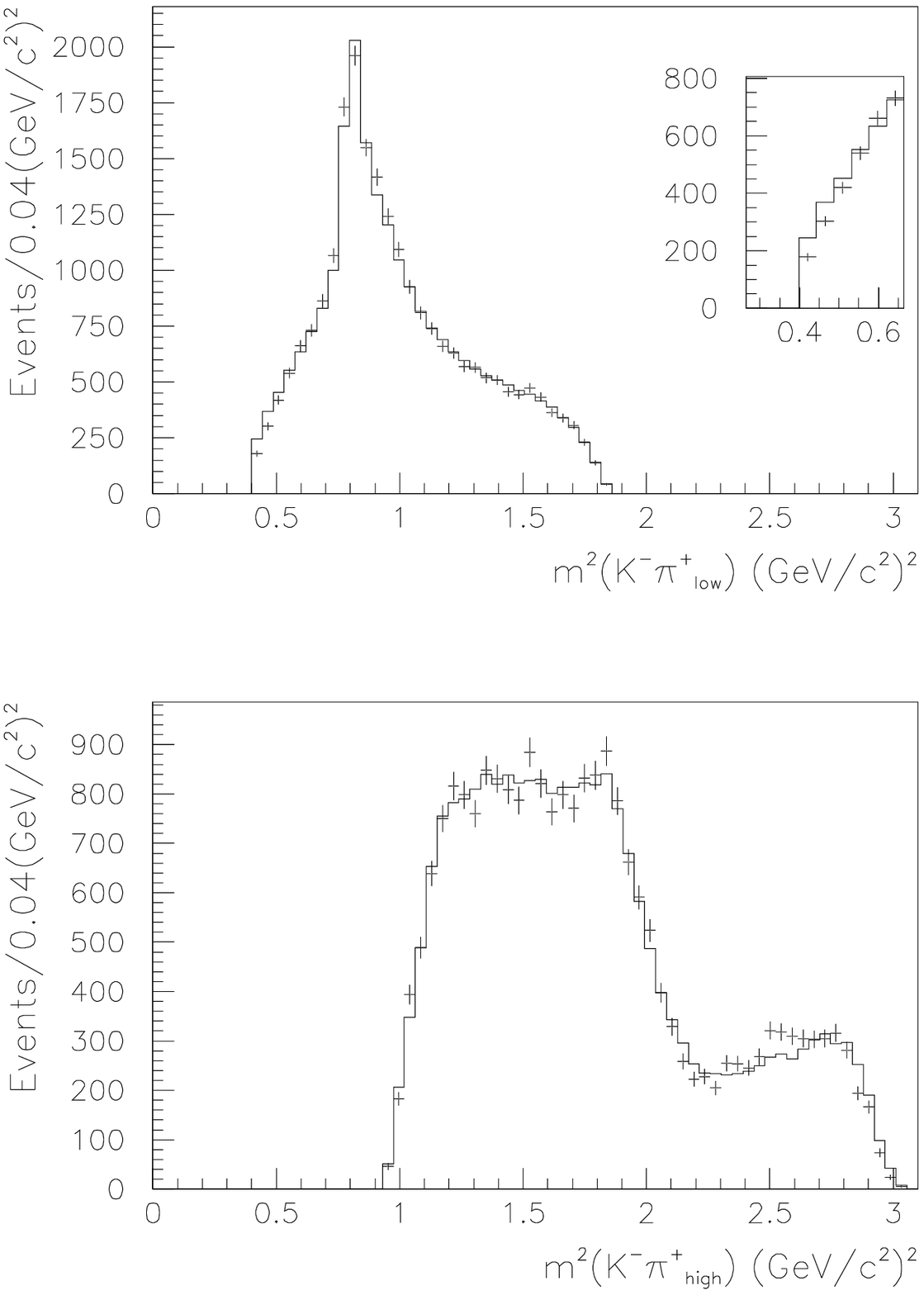}
\includegraphics{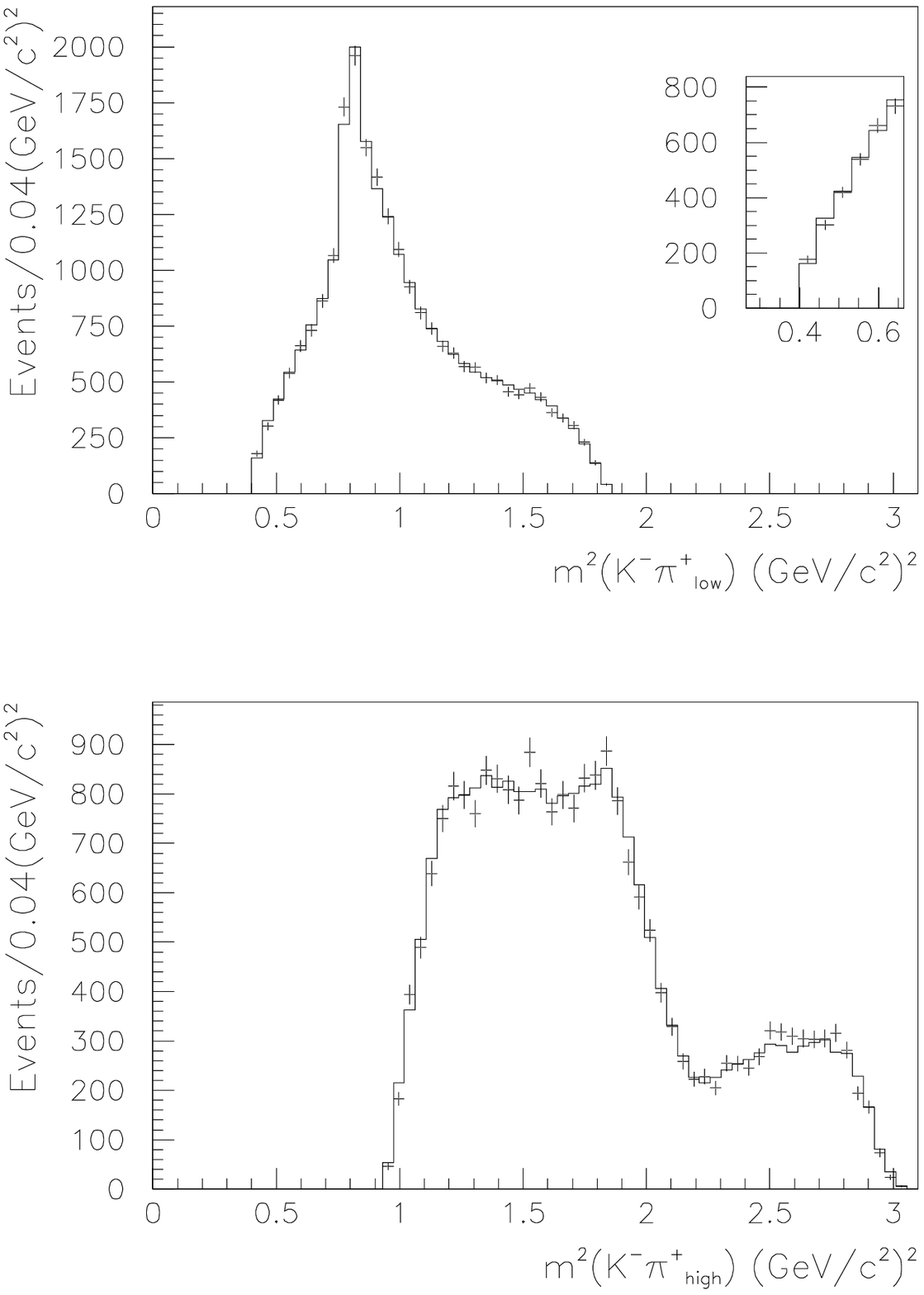}
\hspace*{3cm} (a) \hspace*{5cm} (b)
\caption{\it $m^2(K\pi_{low})$ and $m^2(K\pi_{high})$ projections for \kpipi data (points) and 
our best fit (solid) for models {\rm (a)} without and {\rm (b)} with $\kappa\pi^+$ state. 
\label{proj_kpipi} }
\end{figure}

However, this model does not provide a good description of the data. The calculated
$\chi^2/\nu$ from the binned Dalitz plot distribution is 2.7. 
By observing the $m^2(K\pi_{low})$ and $m^2(K\pi_{high})$ projections for data (points) and
model (solid) in Fig.~\ref{proj_kpipi}(a), we see discrepancies at low $K\pi$ mass squared
(below 0.6 GeV$^2$/c$^4$, where data has very small error bars) and also near 2.5 GeV$^2$/c$^4$.
\begin{table}[t]\centering
\caption{\it Dalitz fit results for \kpipi. {\rm \bf PRELIMINARY}; statistical errors only.}
\vskip 0.1 in
\begin{tabular}{|c|c|c|c|c|} \hline  
 Decay & \multicolumn{2}{|c|}{Fit without $\kappa\pi^+$} 
       & \multicolumn{2}{|c|}{Fit with $\kappa\pi^+$} \\ \cline{2-5}
Mode                      &   Phase($^\circ$)   &    Fraction(\%) 
                          &   Phase($^\circ$)   &    Fraction(\%)  \\ \hline 
non-reson.                & $0$ (fixed)   & $103.9\pm 2.3$ 
                          & $0$ (fixed)   & $ 52.4\pm  8.5$ \\ 
$\kappa\pi^+$             &        --           &         --
                          & $187\pm 11$   & $ 20.7\pm  5.4$ \\ 
${\bar K^*(890)\pi^+}$    & $ 51\pm  1$   & $ 12.7\pm 0.4$ 
                          & $ 15\pm  6$   & $ 11.3\pm  0.4$ \\ 
${\bar K^*_0(1430)\pi^+}$ & $ 63\pm  1$   & $ 33.6\pm 1.4$
                          & $ 63\pm  5$   & $ 17.6\pm  1.7$ \\ 
${\bar K^*_2(1430)\pi^+}$ & $ 51\pm  6$   & $  0.5\pm 0.1$ 
                          & $348\pm 10$   & $  0.3\pm  0.1$ \\ 
${\bar K^*(1680)\pi^+}$   & $ 70\pm  3$   & $  3.8\pm 0.3$ 
                          & $ 53\pm  6$   & $  3.3\pm  0.7$ \\ \hline
\end{tabular}
\label{tabkpipi}
\end{table}

The strong evidence for the $\sigma(500)$ from the \d3pi decay leads us to question whether
a new resonance state could be responsible for the difficulty in modeling the \kpipi data.
The light and broad scalar $K\pi$ resonance, the $\kappa$ (a $\sigma$-meson nonet
member), is presently the subject of many discussions, with controversial results from
$K\pi$ scattering analyses\cite{lass,refkappa}. 

We introduce in our model an extra spin-0 resonant amplitude.
Both mass and width of this hypothetical $\kappa$ state are allowed to float in the 
fitting procedure. Moreover, since the mass and width of the other scalar resonance, $K^*_0(1430)$, 
were obtained from the LASS 
experiment in the absence of this extra amplitude\cite{lass}, we also allow these parameters to 
float. The fit converges to values of the $\kappa$ mass and width of $815\pm 30$ MeV/c$^2$ and
$560\pm 116$ MeV/c$^2$ respectively, and to values for the $K^*_0(1430)$ mass and width of
$1465\pm 6$ MeV/c$^2$ and $182\pm 10$ MeV/c$^2$ respectively. The results for $\kappa$ are consistent
to this state being a light and broad resonance \cite{refkappa}. The results for the
$K^*_0(1430)$ show this state being narrower and heavier than obtained by LASS \cite{pdg,lass}.
The decay fractions and phases from this model are shown in Table \ref{tabkpipi}, second column.
The non-resonant fraction drops to about 50\% and the $\kappa\pi^+$ accounts for about 20\% 
of the decay,
representing the second main contribution, followed by the other scalar state, 
$\bar K^*_0(1430)\pi^+$.
The fit quality of this model is considerably superior to the model without the $\kappa\pi^+$ state:
the $\chi^2/\nu$ is now $0.9$, the confidence level of the fit is 79\%. As can be seen by the
projections in Fig.~\ref{proj_kpipi}(b), there is very good agreement between model and data.

As was done for testing the $\sigma$ state, we test a variety of other models to check their
ability to explain the data. We represent the extra amplitude by vector, tensor and toy 
(B-W with no phase variation) models, allowing mass and width to float. None of these models
reproduced the data as well as the scalar resonance hypothesis.

\section{Conclusions}
Dalitz plot analyses presented here from E791 data point to the importance of
the scalar resonant states in the $3\pi$ and $K\pi\pi$ final states of $D$ mesons.
From the \ds3pi analysis, we find that about 90\% of the decay rate is due to
$f_0\pi^+$ intermediate states. Almost half of the \d3pi decay rate proceeds through
a new scalar state $\sigma(500)\pi^+$. From \kpipi decay, we present
evidence for a new scalar state $\kappa\pi^+$ and, although about 50\% of the decay rate
corresponds to the non-resonant channel, 40\% comes from the scalar states.
We are able to measure the masses and widths of these scalar resonances.
These results illustrate the potential of many-body $D$ meson decays for the study 
of light-meson spectroscopy. 

\end{document}